\documentclass[prl,twocolumn]{revtex4}
\usepackage{graphicx}

\begin{document}

\title{Phase-sensitive Manipulations of Squeezed Vacuum Field in an Optical Parametric Amplifier inside an Optical Cavity }
\author{Jing Zhang$^{1,\dagger }$, Chenguang Ye$^{1}$, Feng Gao$^{1}$, Min
Xiao$^{1,2}$}
\affiliation{$^{1}$The State Key Laboratory of Quantum
Optics and Quantum Optics Devices, Institute of Opto-Electronics,
Shanxi University, Taiyuan 030006, P.R. China} \affiliation{$^{2}$
Department of Physics, University of Arkansas, Fayetteville,
Arkansas 72701, USA}
\begin{abstract}
Squeezed vacuum field can be amplified or deamplified when it is
injected, as the signal beam, into a phase-sensitive optical
parametric amplifier (OPA) inside an optical cavity. The spectral
features of the reflected quantized signal field are controlled by
the relative phase between the injected squeezed vacuum field and
the pump field for the OPA. The experimental results demonstrate
coherent phenomena of OPA in the quantum regime, and show
phase-sensitive manipulations of quantum fluctuations for quantum
information processing.
\end{abstract}

\maketitle

Manipulations of quantum states of light, during their propagation,
storage, and frequency conversion, are essential for quantum
information processing and quantum networking. Recent experiments
have demonstrated the transmission \cite{one}, slowing down
\cite{two,three,four}, and storage and retrieval \cite{five,six} of
squeezed states of light through electromagnetically induced
transparency in multi-level atomic systems \cite{seven}, which are
important to implement quantum network protocol \cite{eight,nine}.
Another important aspect in manipulating quantum fluctuations is the
phase-sensitive amplification and deamplification of the squeezed
states of light, which are needed in quantum information propagation
and communication. An optical parametric amplifier can be used to
amplify non-classical states such as squeezed states and single
photon states. This process, referred to as the "quantum injected
optical parametric amplification", turns out to be particularly
fruitful in the implementations of discrete-variable and
continuous-variable (CV) quantum information processing, such as
optimal quantum cloning machines \cite{nine1,nine2}, optical quantum
U-Not gate \cite{nine3}, bridge between "microscopic" and
"macroscopic" entanglement \cite{nine4}, and CV all-optical quantum
teleportation \cite{nine5}. Bruckmeier, et al experimentally
demonstrated that improved quantum nondemolition measurements could
be realized by injecting amplitude-squeezed light into the meter
input port of an optical parametric amplifier (OPA) \cite{ten1}.
Recently, Agarwal \cite{ten} has theoretically studied the
phase-sensitive responses of quantum states of light through an OPA
inside an optical cavity operated below threshold \cite{eleven}.
Spectral splitting due to quantum interferences between the input
quantum field and the generated down-converted subharmonic field has
been predicted \cite{ten}.

In this Letter, we experimentally demonstrate such quantum
interference phenomena in the phase-sensitive OPA system inside an
optical cavity with an injected squeezed vacuum state. Previously we
had experimentally demonstrated classical interference between the
generated subharmonic field in the OPA system and the injected
coherent signal beam at the subharmonic frequency \cite{twelve}.
Here, we replace the coherent signal beam with a squeezed vacuum
field generated in another OPO operated below threshold (labelled as
OPO, as shown in Fig.1). Since the output field of the subthreshold
OPO is a quadrature squeezed vacuum state, its relative phase to the
amplification (or deamplification) phase of the OPA (Fig.1) has to
be predetermined (by using a homodyne detection setup with a local
oscillator beam) and controlled, which is quite different from the
injected classical coherent signal beam \cite{twelve}. We
investigate the amplification and deamplification of both the
squeezed and unsqueezed quadratures of the input quantum field,
respectively, and show how such phase-sensitive amplifier can be
used to modify and control the input squeezed states of light. The
degree of squeezing can be improved by further deamplifying the
input squeezed quadrature.

Our experimental setup is shown schematically in Fig.1. A
diode-pumped intra-cavity frequency-doubled (continuous-wave ring
Nd:YVO$_{\text{4}}$/KTP) laser is used as the light sources, which
provide about 200 mW of the second-harmonic light at 532 nm and 50
mW of the fundamental light at 1064 nm simultaneously. The
second-harmonic light is divided into two parts to pump the OPO and
the OPA systems respectively. The OPO and OPA systems have the same
structure, each of which has a cavity composed of two coupling
mirrors with same radius of curvature of 30 mm. A 12 mm long PPKTP
(periodically-poled KTP) crystal (Raicol Inc.) with
antireflection-coated flat surfaces for both wavelengths is placed
in each optical cavity. The reflectivities are $99.5\%$ at 1064 nm
and $60\%$ at 532 nm for the input coupler $M1$, which is mounted on
a PZT to adjust the cavity length. The output coupler $M2$ has a
reflectivity of $93\%$ for the OPO at 1064 nm (and $97\%$ for the
OPA), and is a high reflector $(>99\%)$ at 532 nm. The coatings of
the cavity mirrors also provide a second harmonic power build-up
(few times) when the fundamental wavelength is on resonance with the
cavity. The OPO (and the OPA) cavity length is 60 mm. The
temperature of the PPKTP crystal is actively controlled at
millidegree Kelvin level around the operation temperature
(31.3${{}^{\circ }}$C) for optimizing the optical parametric
down-conversion process at the chosen wavelength.

% FIG. 1
%
\begin{figure}
\centerline{
\includegraphics[width=3.3in]{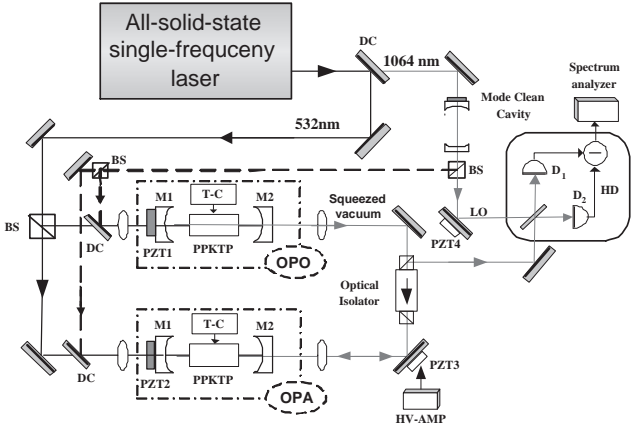}
} \vspace{0.1in}
%\setlength{\columnwidth}{3.2in}
%\centerline{
\caption{ Schematic of the experimental setup for quantum
interference phenomena in the phase-sensitive OPA system inside the
optical cavity with an injected squeezed vacuum state. A squeezed
vacuum state is generated from the subthreshold OPO, then injected
into the OPA as the input signal. DC: dichroic mirror; $\lambda /2$:
half-wave plate; $D_{1},D_{2}$: detectors; T-C: temperature
controller, HV-AMP: high voltage amplifier; PZT: piezoelectric
transducer; BS: beam splitter; LO: local oscillator; HD: homodyne
detection. \label{Fig1} }
%}
\end{figure}

The OPA is pumped by a second harmonic beam with a wavelength of 532
nm through $M1$, and the signal input beam is injected through the
mirror $M2$ from right. The signal field goes through the optical
isolator before entering the OPA cavity. When it goes through the
same optical isolator after reflecting from the OPA cavity, its
polarization rotates 90 degrees from its original polarization axis,
and is reflected by the polarization beam splitter cube. This
reflected signal beam then combines with the local oscillator beam
to be measured by the balanced homodyne detector [12], as shown in
Fig.1. Before injecting the squeezed state into the OPA cavity, we
first characterize the phase-sensitive optical amplifier by
injecting a weak coherent signal beam at the subharmonic wavelength
\cite{twelve}. When the pump beam is blocked, the cavity reflected
signal spectrum is directly detected by a photodetector (not using
the homodyne detection setup), which is shown in Fig.2(a) as a
simple Lorentzian profile. When the pump beam is turned on, but at a
lower power (at 0.2 $P_{th}$, where $P_{th}$ is the OPO threshold
for this OPA+cavity system), the generated subharmonic field is
either in-phase (amplifying) or out-phase (deamplifying) with the
input coherent signal field (by tuning the phase of the injected
signal beam with the mirror mounted on PZT3), as shown in Figs.2(b)
and 2(c), respectively. As the pump power is increased to 0.5
$P_{th}$, the coherent signal is greatly amplified at the resonance
when it is in-phase with the OPA amplifier (Fig.2(d)), but
suppressed (deamplified) when it is out-phase with the OPA
(Fig.2(e)). As one can see that shoulders appear for both amplified
and deamplified spectra just outside the resonant peak (dip), which
indicate that small amplification can occur when the cavity is
detuned from the resonance ( Fig.2(e)) even for the deamplified
operation at resonance. Similarly, small deamplification can appear
in the amplified phase for the OPA outside the resonance (Fig.2(d)).
Notice that the amplified peak (and deamplified dip) has a narrower
linewidth than the input coherent signal dip reflected from the
``empty cavity" (Fig.2(a)) due to such deamplification shoulders in
the amplification phase (and the same is true for the deamplifying
phase with the amplification shoulders).

% FIG. 2
%
\begin{figure}
\centerline{
\includegraphics[width=3.3in]{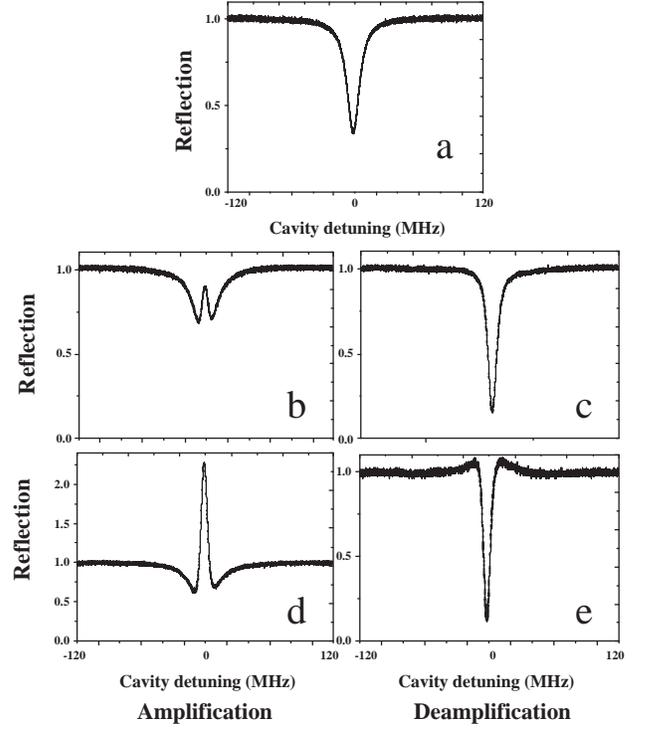}
} \vspace{0.1in}
%\setlength{\columnwidth}{3.2in}
%\centerline{
\caption{ The reflection spectra of the subharmonic field from the
OPA cavity injected with the coherent signal beam, as a function of
the cavity detuning for different pump powers. (a) without the pump
beam; (b) and (c) operated in-phase (amplifying) and out-phase
(deamplifying) at resonance between the injected signal beam and the
pump beam, respectively, with the pump power at 0.2 $P_{th}$; (d)
and (e) with the pump power at 0.5 $P_{th}$. \label{Fig1} }
%}
\end{figure}

A quadrature squeezed vacuum state is generated from the
subthreshold OPO. Due to the high nonlinear coefficient of the PPKTP
crystal, the measured threshold power of the OPO is $P_{th}=80$
$mW$. The OPO correlates the upper and lower quantum sidebands of a
vacuum field that enters the OPO around the center frequency
$\omega_{0}$. The correlation of the quantum sidebands appears as
the squeezed vacuum field \cite{thirteen}. When pumped at frequency
$2\omega_{0}$ and operated at 40 mW below threshold, about 2 dB
squeezing at the sideband frequency of 3.5 MHz is detected, which
passes through the optical isolator and is then injected into the
OPA as the input signal. Since the squeezed vacuum state is elliptic
in phase space, it has to be oriented relative to the amplification
axis of the OPA system carefully. This can be done by tuning PZT3
and checked by detecting it with the local oscillator in the
balanced homodyne detector. The injected squeezing vacuum field is
reflected by the OPA cavity. When passing through the Faraday
rotator again, the polarization plane of the backward light is
rotated $45^{0}$ in the same direction as the initial tilt. This
reflected light is then completely reflected by the polarizer just
behind the Faraday rotator, combined with the local oscillator field
from the laser at a 50/50 beam splitter, and detected by the
balanced homodyne detector system. The local oscillator beam passes
through a mode-cleaner cavity, so that it is in a similar beam
profile as the reflected signal beam from the OPA cavity, and a
fringe visibility of $\sim 96\%$ between them has been achieved in
the experiment.

% FIG. 3
%
\begin{figure}
\centerline{
\includegraphics[width=3.3in]{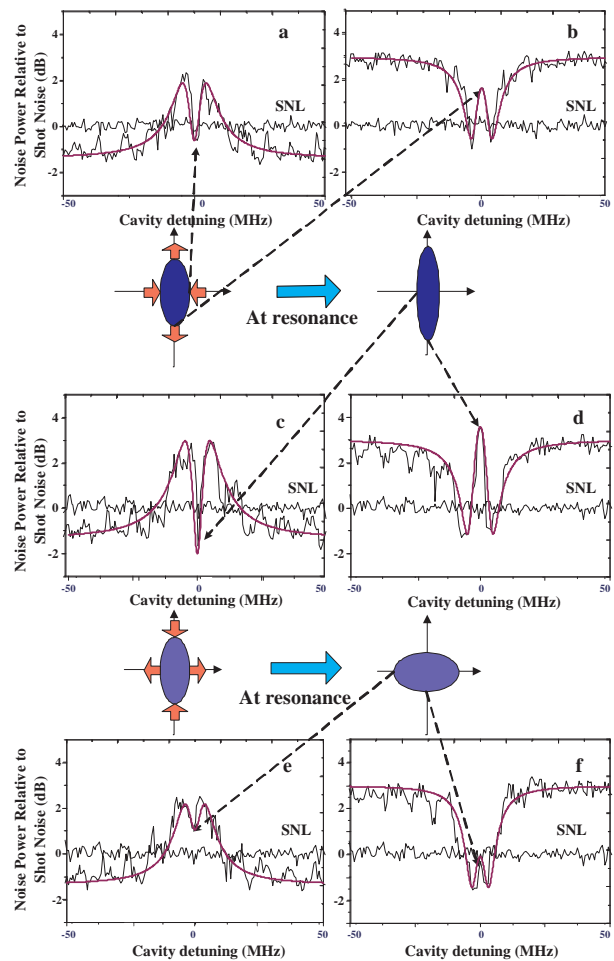}
} \vspace{0.1in}
%\setlength{\columnwidth}{3.2in}
%\centerline{
\caption{(Color online). The reflection spectra of the subharmonic
field from the OPA cavity injected with the squeezed vacuum field as
a function of the cavity detuning. Noise levels are displayed as the
relative power compared to the shot-noise limit. Quantum noise is
measured at the sideband frequency of 3.5 MHz. (a) and (b) without
the pump beam; (c),(d),(e) and (f) with pump power at 0.5 $P_{th}$.
The curves plotted with the purple lines are the theoretical
calculations. \label{Fig1} }
%}
\end{figure}

First, let us look at the squeezed quadrature of the input signal
beam as shown in the diagram below Fig.3(a) (e.g. $\theta=0$) by
choosing the phase of the local oscillator (PZT4) relative to the
input squeezed vacuum state. When the pump beam for the OPA is
blocked, while fixing the phase between the local oscillator and the
reflected vacuum squeezing, and scanning the length of the OPA
cavity, the reflected squeezing signal spectrum is shown in
Fig.3(a). The center part of the reflected spectrum is modified by
the optical cavity with the middle (on resonance) barely below the
shot-noise limit (SNL) (but above the degree of squeezing at far off
resonance), which is used as a reference level for input squeezing.
This spectral shape is induced by the absorption and dispersion
properties of the ``empty cavity" \cite{forteen}. The OPA cavity, as
the empty cavity in this case, is an over-coupled resonator for the
subharmonic field due to $\gamma _{out}>\gamma _{in}+\gamma _l$,
where $\gamma _{l}$,$\gamma _{in}$ and $\gamma _{out}$ are the decay
rates of the subharmonic field resulting from internal losses, the
input mirror M1, and the output mirror M2, respectively
\cite{fifteen}. Next the pump beam for the OPA is turned on. The
relative phase between the injected squeezing signal and the pump
beam of the OPA is tuned by PZT3, which determines the amplification
or deamplification of the signal beam at resonance. When the
relative phase is tuned to be $\phi=\pi/2$, the input squeezing
quadrature is deamplified by the OPA with the degree of squeezing at
the line center to be below the degree of squeezing at far detuning,
as shown in Fig.3(c). This indicates a further squeezing by the OPA
for the initial squeezed quadrature. As the phase of the signal is
tuned in-phase with the pump beam ($\phi=0$), the squeezed
quadrature is amplified, which reduces the initial degree of
squeezing and makes it above the SNL, as shown in Fig.3(e). The
system is operated at a pump power of 0.5 $P_{th}$. The shoulders in
the amplification and deamplification spectra just outside the
resonant dip (or peak), as shown on Figs.2(d) and 2(e), also modify
the reflected spectral shapes at the sides of the peaks in Figs.3(c)
and 3(e).

Now let us examine the situation with the unsqueezed quadrature as
the input signal (e.g. $\theta=\pi/2$). Figure 3(b) shows the
reflected signal spectrum without the pump beam for the OPA. As the
pump beam is turned on and $\phi$ is set at $\phi=\pi/2$, the
deamplifiction for the squeezed quadrature (Fig.3(c)) becomes
amplification for the unsqueezed quadrature, as shown in Fig. 3(d).
The central peak is significantly amplified. As the signal phase is
tuned $90^{0}$ relative to the phase of the pump beam ($\phi=0$),
the amplification for the squeezed quadrature (Fig.3(e)) becomes
deamplification for the unsqueezed quadrature by the OPA (Fig.3(f)),
where the unsqueezed central peak begins to fall below the SNL. The
effect of the amplification shoulders in the deamplification phase
can be seen more clearly here (Fig.3(f)) as the shape of the dip
becomes sharper near the baseline (comparing to the dip in
Fig.3(b)). To make quantitative theoretical comparisons with the
above experimental results, the homodyne spectra for the amplitude
and phase quandratures \cite{sixteen} are calculated with
experimental parameters including all loss mechanisms. Our
theoretically calculated results (purple lines) are plotted together
with the experimental data in Fig. 3, which show excellent
agreements. These theoretical curves can only be qualitatively
compared with the results presented in Ref. \cite{ten}, where power
spectra were given without including any losses.

Figure 3(c) represents a case where the deamplification phase of the
OPA is chosen relative to the squeezed quadrature of the input
squeezed vacuum state. The degree of squeezing has been further
improved below its input level, which can be used as a ``squeezing
amplifier" for applications in quantum communication and quantum
information processing. In principle, the degree of squeezing for
the input squeezed vacuum state can be increased further by
improving the quality of the OPA+cavity system, such as reduction of
internal losses, and operating the system closer to the OPO
threshold, where the quantum nature of the generated field will be
more pronounced \cite{eleven}. There are two significant points
coming out of this experiment: 1) Generating stronger squeezing by
an OPA with an injected squeezed vacuum may improve
Einstein-Podolsky-Rosen (EPR) entanglement further if we use two
squeezed lights and a 50/50 beam splitter to produce the EPR
entangled beam; 2) Injecting other quantum states into an OPA can
lead to important advances in quantum information processing
\cite{nine1,nine2,nine3,nine4,nine5}.

In summary, we have experimentally demonstrated quantum
interferences between the input squeezed vacuum state and the
quantum field generated in the OPA system at different quadrature
phases. We have shown the manipulations of quantum fluctuations due
to quantum interferences by the OPA inside an optical cavity, as
predicted by Agarwal \cite{ten}. Our experiment also indicates that
the input squeezing can be further enhanced to a higher degree when
appropriate relative phase is chosen between the input quantum
signal field and the pump field for the OPA. Such manipulations of
quantum fluctuations by a phase-sensitive optical amplifier (a
quantum state generator itself) are essential in quantum information
processing and quantum networking \cite{seventeen}.

$^{\dagger} $Corresponding author's email address: jzhang74@yahoo.com,
jzhang74@sxu.edu.cn

\smallskip \acknowledgments
J. Zhang thanks K. Peng and C. Xie for the helpful discussions. This
research was supported in part by NSFC for Distinguished Young
Scholars (Grant No. 10725416), National Basic Research Program of
China (Grant No. 2006CB921101), NSFC Project for Excellent Research
Team (Grant No. 60821004), National Natural Science Foundation of
China (Grant No. 60678029), Doctoral Program Foundation of the
Ministry of Education, China (Approval No. 20050108007), the PCSIRT
(Grant No. IRT0516), the TYMIT and TSTIT of Shanxi.

\end{document}